# A RECONFIGURABLE SPINTRONIC DEVICE FOR QUANTUM AND CLASSICAL LOGIC

Debanjan Bhowmik[1,*], Aamod Shanker[2], Angik Sarkar[1], Tarun Kanti Bhattacharyya[2]

[1]Department of Electrical Engineering

[2]Department of Electronics and Electrical Communications Engineering

Indian Institute of Technology, Kharagpur

India - 721302

**ABSTRACT**

*Quantum superposition and entanglement of physical states can be harnessed to solve some problems which are intractable on a classical computer implementing binary logic[1]. Several algorithms have been proposed to utilize the quantum nature of physical states and solve important problems. For example, Shor's quantum algorithm[2] is extremely important in the field of cryptography since it factors large numbers exponentially faster than any known classical algorithm. Another celebrated example is the Grover's quantum algorithm[2], which provides quadratic speedup in searching an unsorted database. These algorithms can only be implemented on a quantum computer which operates on quantum bits (qubits). Rudimentary implementations of quantum processor have already been achieved through linear optical components[3], ion traps [4], NMR [5] etc. However demonstration of a solid state quantum processor had been elusive till DiCarlo et al demonstrated two qubit algorithms in superconducting quantum processor[6]. Though this has been a significant step, scalable semiconductor based room temperature quantum computing is yet to be found. Such a technology could benefit from the vast experience of the semiconductor industry. Hence, here we present a reconfigurable semiconductor quantum logic device (SQuaLD) which operates on the position and spin degree of freedom of the electrons in the device. Based on a few recent experiments, we believe SQuaLD is experimentally feasible. Moreover, using a well known quantum simulation method, we show that quantum algorithms (such as Deutsch Jozsa, Grover search) as well as universal classical logic operations (such as NAND gate) can be implemented in SQuaLD. Thus, we argue that SQuaLD is a strong candidate for the future quantum logic processor since it also satisfies the DiVincenzo[7] criteria for quantum logic application as well as the five essential characteristics for classical logic applications[8].*

In the last decade, many different implementations of quantum logic have been proposed in semiconductors based systems. Most of these involve spin qubits where information is coded by spin degrees of freedom. This is primarily because spin is believed to be much more robust and stable than charge qubit and has longer lifetime in most semiconductor systems. However, most of the well-known spin qubit based proposals involve stationary qubits where the physical system encoding the qubit is fixed in space. Examples in this category include the proposal to manipulate nuclear spins [9] and

- Corresponding author email debanjan@eecs.berkeley.edu  ( has currently moved to Department of Electrical Engineering and Computer Science at University of California , Berkeley) electron spins in quantum dots [10,11]. Another class of spin qubits that have recently garnered attention are the flying qubits where, contrary to the previous case, quantum gates are fixed, while the position of the qubits changes with time. SQuaLD also implements quantum operations on flying qubits. However, the quantum gates in our device are not entirely fixed and can be altered by changing an external tuning voltage. SQuaLD also has certain other features in common with previous work. For example, as in previous proposals[12,13], it uses parallel semiconductor nanowires . However, ref. 12 requires local magnetic fields which are technologically challenging and power consuming. We avoid magnetic fields at all stages in our device operation. Also these references assume that a junction of two nanowires is a beam splitter which we have found to be inaccurate through our analysis. So we present a theoretical model of how such beam splitting action can be carried out on ballistic electrons through the use of potential barrier in Appendix A of Supplementary Section. To the best of our knowledge there has been no theoretical model for such device developed so far.Moreover unlike these proposals, we envisage universal quantum operations to be possible for all qubits in our device. This flexibility is extremely important in implementing various complex quantum logic operations. In short, SQuaLD can implement all quantum and classical logic operations enabling us to demonstrate for the first time, quantum algorithms like Grover search and Deutsch Jozsa in a semiconductor based device.

Implementation of classical logic on a quantum logic system may seem trivial since a classical bit can be thought of as the basis state of a qubit. Indeed, almost any classical logic gate can in theory be realized by a sequence of quantum logic gates which implement the reversible universal gate (e.g. the Toffoli gate)[2]. However, it is difficult to realize a device that would simultaneously satisfy all the necessary criteria for quantum as well as classical logic applications. Nevertheless, it is imperative to implement both quantum and classical logic in the same device for the future quantum logic processor. This is primarily because of the non-deterministic nature of the most quantum algorithms. For example, Shor's algorithm for factoring large numbers has a definite probability of yielding erroneous results. So the results have to be checked and the algorithm rerun if the answers were wrong. In this case, the veracity of the results can be verified with much lesser resources in a classical computer which multiplies the factors. Thus, implementation of classical logic in the quantum logic device is extremely advantageous from the point of view of the future quantum processor, which we achieve in this proposal.

The basic unit of SQuaLD consists of a pair of ballistic nanowires (Fig 1). Electrons are injected into the nanowires and various gates manipulate the state of these 'flying' electrons. The quantum state of an electron in the nanowire can be described by a two qubit wavefunction, $|\Psi> = |k, \sigma>$. 'k' refers to the "which nanowire" degree of freedom (|0> or |1>) of the electron with localization in either of the nanowires being its basis states. Following the terminology used to describe charge localization in bilayer graphene [14] we refer to this degree of freedom as the 'pseudo spin' qubit. 'σ' refers to spin qubit with 'up' spin ($|\uparrow>$) and 'down' spin ($|\downarrow>$) being its basis states. In Fig 1, the "coupling" gate connecting two adjacent unit cells of SQuaLD operates exclusively on the pseudo-spin qubit by redistributing the electrons coming from either or both of the input nanowires into the output nanowires with an unique control over their phase relationship. A classic example of such device is the mesoscopic electron beam splitter [15].In Appendix A , we derive a theoretical model for it.On the other hand, the "field" gate performs spin precession using Rashba effect[16], thus manipulating the spin qubit. The amount of spin precession determines the gate operation on the spin qubit and can be controlled by the voltage applied to the field gate which creates the transverse electric field needed for spin-orbit interaction (Rashba effect). This makes SQuaLD reconfigurable since different gates can be implemented in the same device configuration just by changing the voltage applied to the field gate.

Experimental efforts over the last 15 years lead us to believe that SQuaLD is experimentally feasible. For example, gate control of spin precession or spin qubit manipulation has recently been experimentally demonstrated in InAs heterostructure [16,17]. The beam splitter being experimentally demonstrated by Liu et al [15], other "coupling " gates are also experimentally feasible. This is because other "coupling" gates can be obtained by combining the beam splitter with phase shifters[2]. To implement a phase shifer, i.e. to introduce a path difference between the electrons in the two parallel nanowires we have to insert a potential barrier in one of them.

In this paper, we use Non Equilibrium Green's Function (NEGF) based quantum simulation method to simulate electron transport and show quantum and classical logic operation in SQuaLD. The simulation model is explained in Appendix B of Supplementary Section. To illustrate quantum logic operations, we present the implementation of Deutsch-Jozsa (DJ) (Fig 2) and Grover search algorithm (Fig 3) in SQuaLD. DJ is a quantum algorithm that can determine whether a function mapping from N bits to 1 bit is a constant function(always 0(1)) or a balanced function (1 and 0 equally probable). Classically such an evaluation would require $2^{N-1}+1$ function calls. However DJ is impressive since it performs the evaluation in just one function call [2] irrespective of the value of N. DJ can be well illustrated by a simple case – function mapping from one bit to one bit. There are four possible functions, constant functions: $f_0(x)=0$, $f_1(x)=1$ x={0,1} and balanced functions $f_2(x)=x$, $f_3(x)=1-x$ where x={0,1}. Classically the evaluation would have taken at least two function calls, one call with each possible value of x= {0,1}. However we have shown in Fig 2 that such an evaluation can be made in one call using DJ in SQuaLD. We implemented DJ in SQuaLD using the quantum circuit with the gate sequence shown in Fig 2a. In our simulation scheme, pseudo-spin is the main qubit and spin of electron is the ancillary qubit. We initialize the qubits to $|0\downarrow>$ (down spin electrons with pseudo-spin '0') and pass it through the sequence of gates (explained in details in Appendix C of Supplementary Information). The algorithm has been tested for 4 different cases of the oracle implementing $f_0(x)$, $f_1(x)$, $f_2(x)$, $f_3(x)$ (Fig 2b). For each of the 4 different cases of the functions ($f_0(x)$, $f_1(x)$, $f_2(x)$, $f_3(x)$) implemented by the oracle, SQuaLD correctly indicates if the function is constant or balanced, as demonstrated by the NEGF simulation signatures (Fig 2c)

DJ illustrates the power of quantum computing. However it is not very useful in a practical application. So, here we also present simulations of Grover search algorithm, one of the most famous quantum algorithms having a practical use. It searches an unsorted database with N entries in $O(N^{1/2})$ time and using $O(\log N)$ storage space. An equivalent classical algorithm would be able to complete the task only in linear time. We implement a two qubit version of Grover algorithm, where the required string is searched from 4 strings in just one function call. This has been achieved by the gate sequence diagram shown in Fig 3a with spin and pseudo-spin as the two qubits . For simplicity, we assume that the oracle marks searched string with a negative sign .The marked string has been correctly identified by SQuaLD (Fig 3).

The relevance of performing classical computation in a quantum logic device has already been discussed. We now present simulations of the universal NAND gate (Fig 4) which can be used in different combinations to implement any classical logic gate. The gate sequence diagram of NAND gate consists of the SQuaLD with beam splitter at the two ends. The two beam splitters implement a NOT gate on the pseudo-spin qubit, without affecting the spin of the electron. Such beam splitters have already been fabricated. Thus, using this property of beam splitters, we implement NAND gate in SQuaLD with spin (|↑>:'0';|↓>:'1') and pseudo-spin ('0' or '1') inputs. For output, we envisage that an up-spin electron (↑:'0') detector would be coupled to the pseudo-spin '1' channel at the output end so that the output is '1' (detected) when the input state is |1↑> and '0' in all other cases. Thus we get the truth table of NAND gate as shown in Fig 4.

We believe QLD will restrictively satisfy the five celebrated Di Vincenzo criteria for physical realization of a quantum computer:

1. Well defined qubits and scalable physical system-Spin and pseudo-spin form an ideal two qubit system .Spin and pseudo spin both can exist in only the two basis states and superposition of these states. Moreover they can also exist in entangled states – for example, the states created by operation of CNOT gate on a pure state. When the system is scaled up, a N qubit wave function in our device can be represented as where $I^{N-1}$ represents N-1 dimensional identity matrix. Physically it translates to having N-1 pairs of nanowires. The number of gates required in such a manipulation can be reduced significantly by following the protocol established by in similar optical interferometric systems[19]. For algorithms that do not require entanglements, N qubits can also be implemented by N/2 spins in N/2 pairs of nanowires.
2. Ability to initialize state of qubits to simple fiducial state - The advantage of SQuaLD is that it has an all electrical interface i.e. states can be initialized by regulation of electrical voltages only. The spin qubit can be initialized to |↑> or |↓> electrically by two methods. The first one involves passing unpolarized current through the mesoscopic Stern-Gerlach apparatus (MSGA) proposed by Ionociou [20]et.al.. The MSGA exploits a structure similar to our device; but involves local magnetic fields. The second method involves using a ferromagnetic contact which acts as a spin filter. In our proposal, we use the MSGA for two reasons. Firstly, the MSGA structure is similar to ours, so it can be easily coupled to our device. Secondly, MSGA and SQuaLD can be made in the same material eliminating the interface related spin injection issues.
    Injection of the pseudo-spin too can be regulated electrically by manipulation the voltage controlled Rashba field induced by the Field gate. Let us consider an unit cell of QLD with beam splitter at the two ends. When the field gate is turned off and the Rashba field is absent, the pair of beam splitters acts as inverter. Electrons are inserted from input port '0' as in Figure 1. Electrons will be go to output port '1' ie their pseudo spin qubit will become |1>.On the other hand, when field gate is turned on with the Rashba field at 1x10$^{-10}$ eV/m, the electrons go to output port '0' ie their pseudo spin qubit will become |0>. (We prove this analytically and through simulations in Appendix D of Supplementary Section)Thus pseudo-spin states can be initialized to |0> or |1> electrically by turning on or off the field gate, this particular aspect is

also useful for classical logic applications as the device will have fully electrical interface for the inputs.

3. Long relevant decoherence times, much longer than operation times- Spin and phase decoherence times are both important in our case. Since our device is based on flying qubits (qubits which move between the input and output), it is more pertinent to refer to decoherence lengths. The spin and phase coherence lengths have been reported to be 3 microns and 300 nm respectively at around 2K.[21,22] Our device dimensions are much smaller than these lengths. Thus, the operation can be safely assumed to be coherent.

4. A universal set of quantum gates- Any universal single qubit operation on the spin qubit can be performed by arbitrary rotations and any such arbitrary rotation can be broken down into successive rotations about the z,y and z axes respectively [2], and such rotation can be brought about by application of transverse electric field causing Rashba effect through the field gate. The phase gates can be used to perform single qubit operations on pseudo spin qubit. But universal quantum operation necessitates operation of a two qubit gate like CNOT along with universal single qubit operations. Implementation of CNOT gate is trivial in our device- presence of a Rashba electric field in the lower electron channel set at such value that it will invert the spin of the electron ( |0> to |1> and |1> to |0>) implements the CNOT gate. Implementation of all these gates on spin and pseudo spin qubits has been shown through NEGF simulations of the quantum algorithms.

5. A qubit specific measuring capability- Standard spin sensors which convert spin to voltage can be used for detecting the spin qubit. The MSGA can also be used for the purpose as it uses a similar structure as ours. If the spin state is to be stored for future use, it may also be stored in the magnetization state of a ferromagnet. The pseudo-spin state can be easily determined by the amount of current flowing through the two channels.

We believe that, four of the five essential characteristics for classical logic applications should be trivially ensured in our device. A 'complete set of Boolean logic gates' can be easily implemented based on the NAND gate that we have already discussed. Since the readout of the gates is via the ferromagnet or voltage based scheme described earlier, the criteria of 'logic Level restoration' is also valid. These readout schemes provide the 'gain' required for the operation. Thus input is isolated from the power supply. The input and output are both in the same form, namely spin/pseudo-spin orientation which can be converted to voltages whenever required. Hence the criterion of 'concatenability' is also ensured.

**Figure Captions**

**Figure 1. A schematic of the unit cell of the Quantum Logic Device(QLD).** It consists of nanowires in which electrons are localized. Localization in either nanowire is the basis state for the pseudo-spin qubit('0' or '1') while spin of the electron forms the second qubit of our two qubit system. The Phase Gate operates on the pseudo-spin qubit only while the Field Gate subjects the moving electron spins to transverse electric field thus giving rise to various spin qubit gates by spin precession owing to Rashba effect. The shading in the nanowire depicts the path of the electron spin in the particular case of field gate turned off and coupling gate being a beam splitter. Since product of two beam splitters is an inverter matrix, pseudo –spin qubit will be inverted ,ie, electrons coming from input port 0( pseduo spin state '0') will go to output port '1'( pseudo spin state '1')

**Figure 2-Deutsch Jotza Algorithm-**a.Gate sequence diagram showing two qubit version of Deutsh Jotza algorithm b.Implementation of the algorithm in SQuaLD is shown, pseudo spin being the main qubit and spin being the ancillary qubit.Hadamard gate on spin is implemented by application of electric field causing Rashba effect at $0.5 \times 10^{-10}$eV/m which rotates the spin of electron ( initially –z polarized) by 90 degrees to make it –x polarized .Implementation of oracle is expalined in details in Appendix C of the Supplementary Section. c. NEGF simulation results verify Deutsch's algorithm in our device. When function is constant ( f(0)=0,f(1)=0 and f(0)=1,f(1)=1 cases) the final state of the main qubit(pseudo spin) should be |0>, so current is high when right contact is made at output port 0 of the device and very low when contact is made at output port 1. Similarly for the other two cases of the function (balanced) final pseudo spin state is |1>, hence current is high when right conact is at output port 1 and very low at output port 0.

**Figure 3-Demonstration of Grover's Search Algorithm-** a.Gate sequence diagram showing two qubit version of Deutsh Jotza algorithm( Number of searched strings is 4) b. Implementation of the algorithm in SQuaLD is shown, pseudo spin being the first qubit and spin being the second qubit.Hadamard gate on spin is implemented by application of electric field causing Rashba effect at $0.5 \times 10^{-10}$eV/m which rotates the spin of electron ( initially +z polarized) by 90 degrees to make it +x polarized c. NEGF simulation results verify Grover's algorithm in our device. Table 1 shows the representation in the two qubits corresponding to the 4 searched strings,for each of the searched strings its corresponding output is obtained e.g for first searched string (|00>) spin up electrons are obtained at output port 0.The results predicted by Table1 match with our simulation results.

**Figure 4-Demonstration of NAND gate- a.** Spin up ('0') or down('1') electrons are introduced from either output port  0 (pseudo spin '0')or 1 (pseudo spin '1'), with a spin detector/filter at output port 1.Since field gate is turned off and each phase gate is a beam splitter, couple of beam splitters acts as inverter. So pseudo spin state is inverted keeping spin the same. So the detector will detect,or output will be'1'(detection occurs) if up spin electrons are introduced from ouput port 0 only, thereby satisfying NAND gate truth table **b.** If the up spin detector is replaced by an up spin filter at output port 1 and right contact is made at that port, instead of detection occurring, we would say that current flow is high. The NEGF simulations show that current is high only when input port is 0 and spin of input electrons is up(0) as expected.

**Figures**

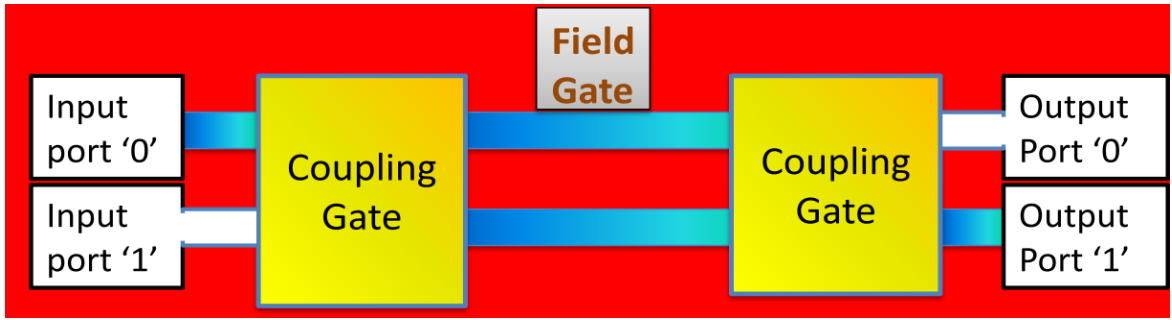

Figure 1

## a

Main qubit

|0⟩ —[H gate 1]— |x⟩ —[Oracle]— |x⟩ —[H gate 3]— [Measurement]

|1⟩ —[H gate 2]— |y⟩ —[Oracle]— |y XOR f(x)⟩

Ancillary qubit

## b

Input port '0' → [Field gate - H2] ← |0⟩ [H gate 1] |0⟩ [Oracle $U_f$] |0⟩ [H gate 3] |0⟩ → Output Port '0'

Input port '1' — |1⟩ — |1⟩ — |1⟩ — |1⟩ → Output Port '1'

Input port '0' → [Field gate - H2] ← |0⟩ [H gate 1] |0⟩ [Oracle $U_f$] |0⟩ [H gate 3] |0⟩ → Output Port '0'

Input port '1' — |1⟩ — |1⟩ — |1⟩ — |1⟩ → Output Port '1'

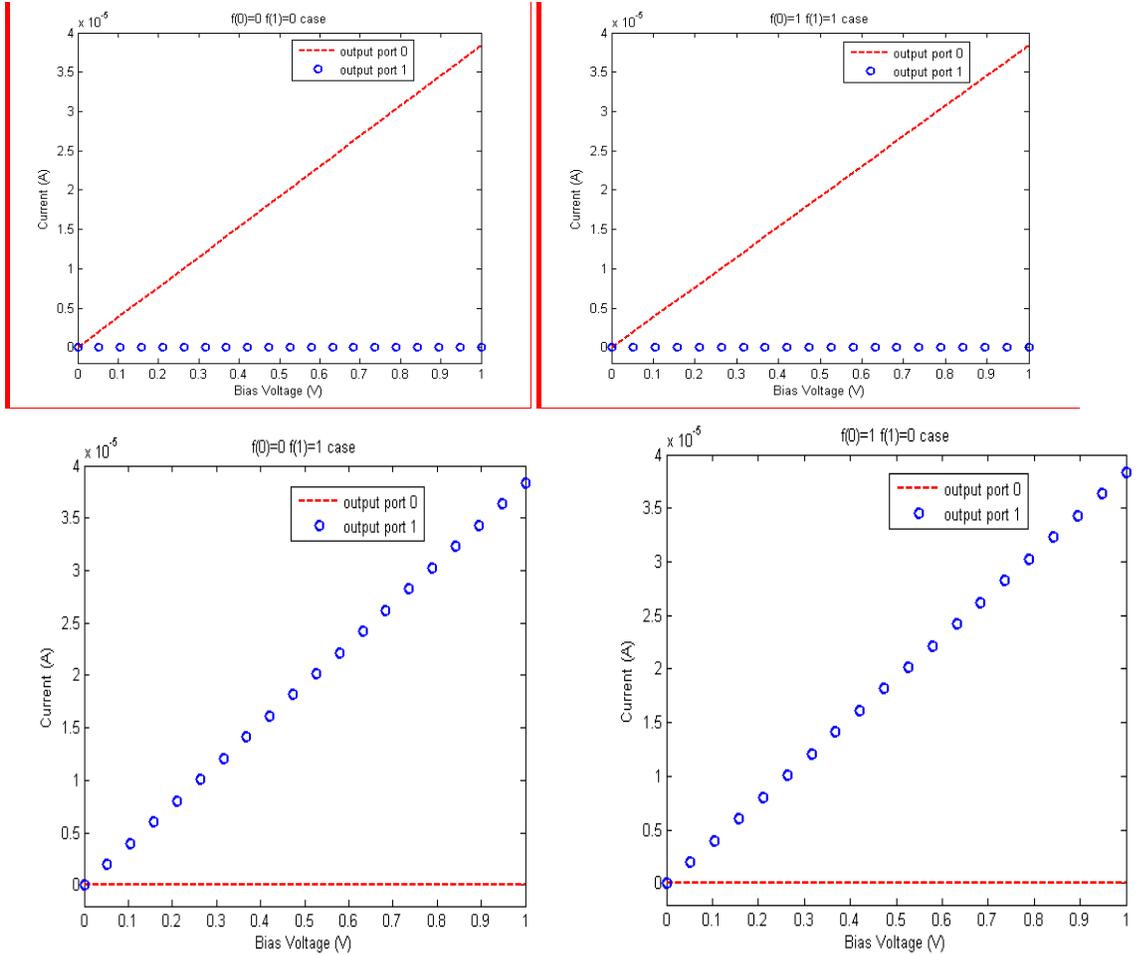

**c**

Figure 2

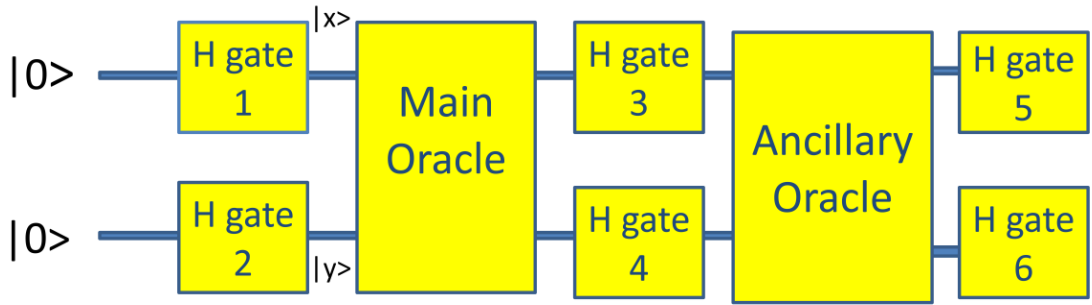

a

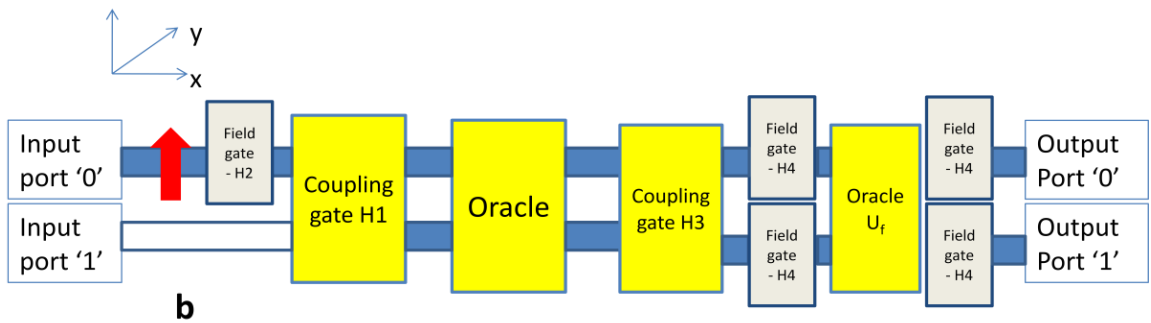

b

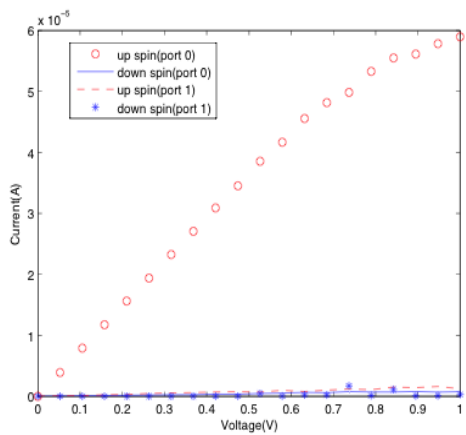
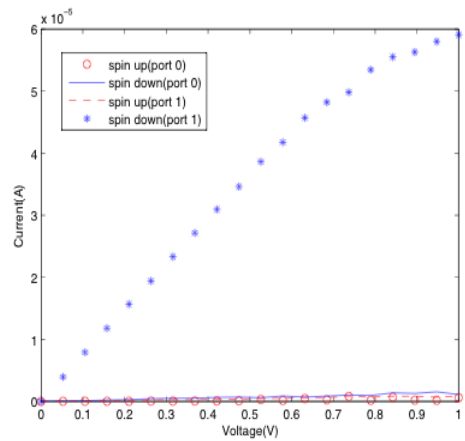

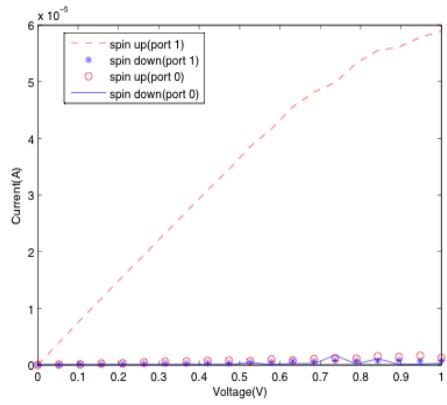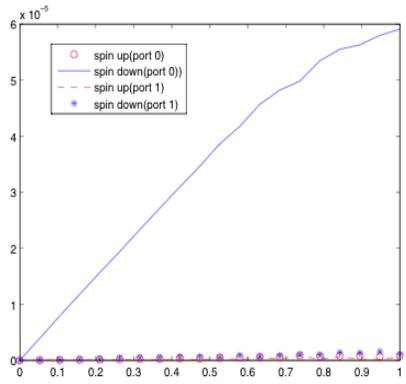

c

Figure 3

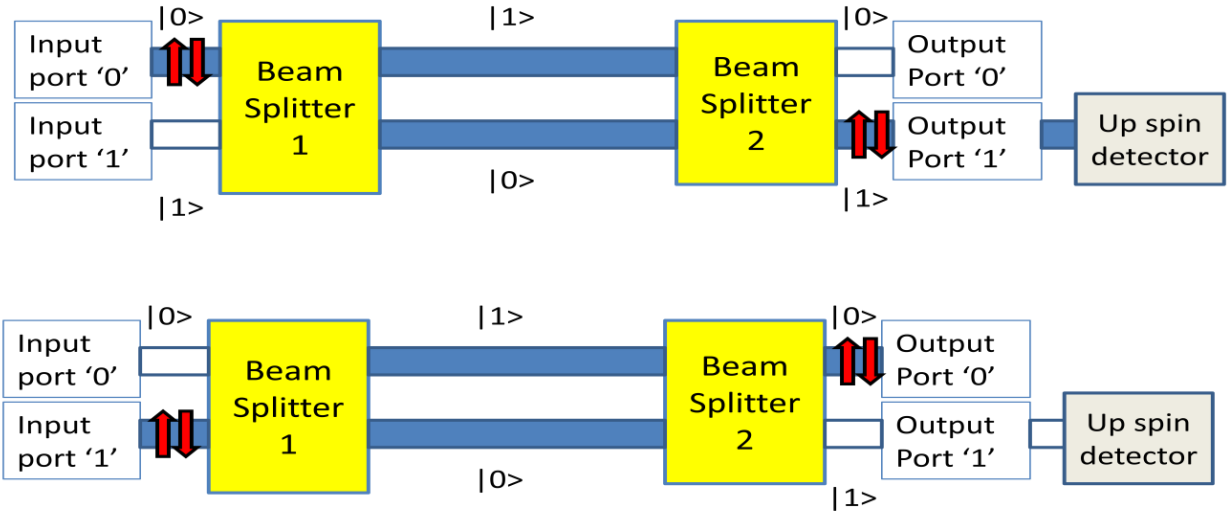

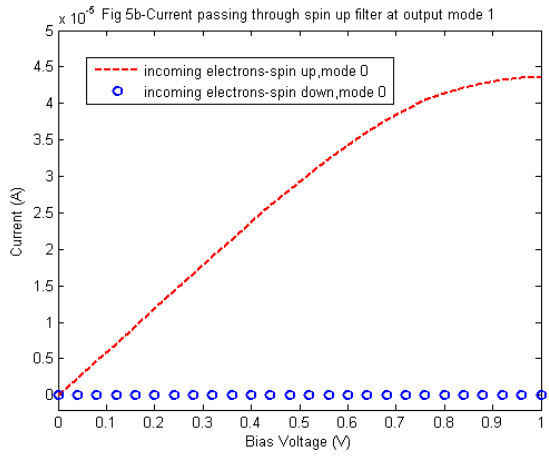

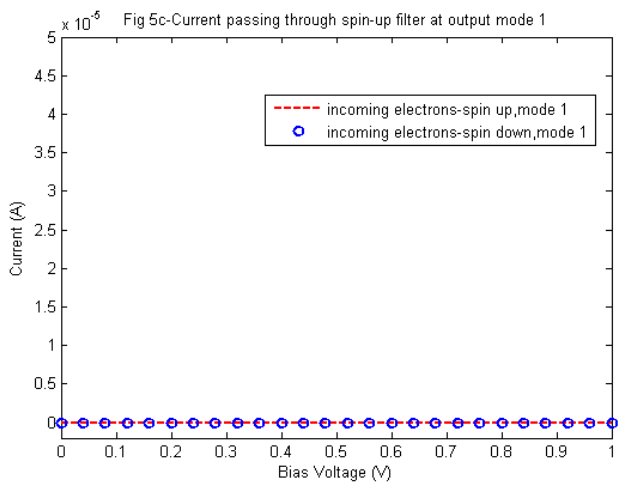

Figure 4

| String | output port | spin |
|---|---|---|
| 1  |00> | 0 | up |
| 2  |01> | 0 | down |
| 3  |10> | 1 | up |
| 4  |11> | 1 | down |

**Table 1-** Corresponding to each string searched in the Grover's algorithm , we expect a particular spin state at a particular port in the output of our device (Figure 3b) which is stated in this table.

# SUPPLEMENTARY SECTION TO ACCOMPANY "A RECONFIGURABLE SPINTRONIC DEVICE FOR QUANTUM AND CLASSICAL LOGIC"

## A. FORMULATION OF THE BEAM SPLITTER

Electron beam splitters are a crucial component of spintronic based reconfigurable Quantum Logical Devices (QLDs). Various experimental groups have been working on constructing a mesoscopic electron beam splitter. It was demonstrated by RC Liu et al. (R.C.Liu, 1998) that a potential barrier of a particular thickness could act like a beam splitter for fermions. To test the beam splitting, an electron wave was made incident on the potential barrier at an angle, and two receivers were placed on either side. Due to the antibunching of the electron, a suppression of the shot noise was expected at the receivers, which was confirmed. However the beam splitter scattering matrix was assumed to be $\frac{1}{\sqrt{2}}\begin{pmatrix} 1 & 1 \\ -1 & 1 \end{pmatrix}$ during the experiment. This was keeping in view the facts that the beam splitter was a 50-50 beam splitter, and that the scattering matrix was unitary (which is directly obtained from the power conservation condition). Hence a -1 was directly assumed to be the third term of the scattering matrix to make the scattering matrix unitary, without a thorough analysis. In the following section, we undertake a rigorous analytical formulation of the beam splitter. We begin with the optical beam splitter, where the solution of the Maxwell's equations (which gives us the Helmholtz equation in E or H) along with the boundary conditions gives us the beam splitting condition and the corresponding scattering matrix. The Schrodinger's equation for a single fermion is also a second order differential equation, much like the Helmholtz equation. We use the similarity of the equations to draw parallel solutions for the electron wave corresponding to the electromagnetic wave. We shall start from the single boundary case and use the results obtained there to the two boundary case. The two boundary structure is expected to give us beam splitting and antibunching of the incident fermions. We shall that the scattering is different from that assumed by RC Liu et al, and it successfully explains the antibunching of the electron wave as observed in the experiment.

**Equivalence of Helmholtz and Schrodinger equations**

The scalar Helmholtz wave equation (obtained from the Maxwell's equations)(Griffith) is given by

$$\nabla^2 \tilde{E} + k^2 \tilde{E} = 0 \qquad (1)$$

with $k^2 = \omega^2 \mu \epsilon$, where equation (1) is the Helmholtz equation for the phasor electric field $\tilde{E}$, $k$ is the propagation constant, $\omega$ is the angular frequency, and $\mu, \epsilon$ are the permeability and permittivity of the medium respectively. The Schrodinger equation for a particle in potential V assuming negligible effective mass ($m^*$) variation in space, is given by

$$\nabla^2 \tilde{\psi} + \frac{2m^*(E - V)}{\hbar^2} \tilde{\psi} = 0 \qquad (2)$$

where $\tilde{\psi}$ is the complex electron wave function, $E (= \hbar\omega)$ is the electron energy and $V$ is the applied potential . Drawing a direct analogy between equations (1) and (2), we can write

$$\nabla^2\tilde{\psi} + \gamma^2\tilde{\psi} = 0 \tag{3}$$

where $\gamma^2 = \frac{2m^*(E-V)}{\hbar^2}$ and represents, as in the Helmholtz equation, the square of the propagation constant. Hence we can clearly draw a parallel between the scalar Helmholtz equation and the Schrodinger equation. For the Helmholtz equation, we can further write $k = \omega\sqrt{\mu\epsilon} = k_0\sqrt{\mu_r\epsilon_r} = k_0 n$ where $n$ is the refractive index of the medium with respect to a reference medium, and $\mu_r, \epsilon_r$ are the relative permeability and permittivity respectively of the medium with respect to the same reference. To treat the electron wave similarly, we can write $\gamma = \gamma_0 n_q$ with

$$n_q = \sqrt{m_r^*(E-V)_r} \tag{4}$$

$m_r^*$ and $(E-V)_r$ are the relative effective mass and the relative energy difference with respect to an arbitrary reference (Daniela Dragoman, 2004). Hence changing the applied potential in a medium for an electron wave effectively changes the "refractive index" of the electron wave in the medium. Analogies for optical structures for EM waves can consequently be derived for electron waves.

**Electron wave incident on a potential step**

The transmission and reflection probability currents for an electron incident on a potential step at an angle can be found from the boundary conditions. The electron wave is assumed  be incident at an angle, so that the input and output port can be distinguished by their spatial separation .Here, the results for transmission and reflection coefficients for an EM wave incident on a dielectric interface are used directly to find the analogous result for the electron wave.

First we consider the solution for an EM wave (Fig S1a). The wave is TEM polarized, with the electric field parallel to the dielectric interface and the magnetic field perpendicular to the electric field as well as the direction of propagation. We assume that $E_i = \tilde{E}_i e^{j(\omega t - k_i.r)}$, $E_r = \tilde{E}_r e^{j(\omega t - k_r.r)}$ and $E_t = \tilde{E}_t e^{j(\omega t - k_t.r)}$ are the incident, reflected and transmitted waves respectively. The phase matching condition at the interface gives us the law of reflection ($\theta_i = \theta_r$) and refraction ($k_i \sin\theta_i = k_t \sin\theta_t$)(Griffith). There are four boundary conditions on the complex amplitudes $\tilde{E}_i$, $\tilde{E}_r$ and $\tilde{E}_t$ , one each corresponding to perpendicular and parallel components of the electric field, and perpendicular and parallel components of the magnetic field. As the electric field has no component perpendicular to the interface, one of the boundary conditions is eliminated. We obtain(Griffith)

Parallel  E field: $\quad\quad\quad\quad\quad\quad \tilde{E}_i + \tilde{E}_r = \tilde{E}_t \tag{5}$

Parallel  H field: $\quad\quad\quad\quad\quad\quad \dfrac{\tilde{E}_i}{Z_1}\cos\theta_1 - \dfrac{\tilde{E}_r}{Z_1}\cos\theta_1 = \dfrac{\tilde{E}_t}{Z_2}\cos\theta_2 \tag{6}$

Perpendicular H field:

$$\mu_1 \left( \frac{\tilde{E}_i}{Z_1} \sin\theta_1 - \frac{\tilde{E}_r}{Z_1} \sin\theta_1 \right) = \mu_2 \frac{\tilde{E}_t}{Z_2} \sin\theta_2 \qquad (7)$$

where $Z_1 = \sqrt{\frac{\mu_1}{\varepsilon_1}}$ and $Z_2 = \sqrt{\frac{\mu_2}{\varepsilon_2}}$ are the $\tilde{E}/\tilde{H}$ ratios of the first and second media respectively. Equation (7) can be derived from equations (5), (6), and from the Snell's law of refraction. Hence we fundamentally have two equations (5) and (6), that can be solved to get $\tilde{E}_i$ and $\tilde{E}_r$ in terms of $\tilde{E}_t$ as follows(Griffith)

$$\frac{\tilde{E}_r}{\tilde{E}_i} = \frac{Z_2 \cos\theta_1 - Z_1 \cos\theta_2}{Z_2 \cos\theta_1 + Z_1 \cos\theta_2} \qquad (8)$$

$$\frac{\tilde{E}_t}{\tilde{E}_i} = \frac{2 Z_2 \cos\theta_1}{Z_2 \cos\theta_1 + Z_1 \cos\theta_2} \qquad (9)$$

For the electron wave, a similar analysis can be performed. In this case the electron wave is incident at an angle on a potential step (Fig S1b), which can be equated to the interface between two dielectric media by the effective refractive index expression (4).

We assume that $\psi_i = \tilde{\psi}_i e^{j(\omega t - \gamma_i \cdot r)}$, $\psi_r = \tilde{\psi}_r e^{j(\omega t - \gamma_r \cdot r)}$ and $\psi_t = \tilde{\psi}_t e^{j(\omega t - \gamma_t \cdot r)}$ are the incident, reflected and transmitted waves respectively. Equating the phases at the boundary, we get the law of reflection and Snell's law for the electron wave. The conserved quantities in this case are the complex wavefunction amplitude $\tilde{\psi}$ (continuity of wavefunction at boundary) and the component of the first derivative perpendicular to the interface $(\nabla. \psi).\hat{z}$ (differentiability of wavefunction at boundary). The conditions can hence be written as :

$$\tilde{\psi}_i + \tilde{\psi}_r = \tilde{\psi}_t \qquad (10)$$
$$(\nabla.\psi_i)\cos\theta_1 - (\nabla.\psi_r)\cos\theta_1 = (\nabla.\psi_t)\cos\theta_2 \qquad (11)$$

To draw an analogy with equations (8) and (9), we introduce the quantity $Y$ (corresponding to intrinsic impedance),

$$Y = \frac{\psi}{|\nabla.\psi|} = \frac{1}{\gamma} = \sqrt{\frac{\hbar^2}{2m^*(E-V)}} \qquad (12)$$

We can hence rewrite equation (11) as:

$$\frac{\tilde{\psi}_i}{Y_1}\cos\theta_1 - \frac{\tilde{\psi}_r}{Y_1}\cos\theta_1 = \frac{\tilde{\psi}_t}{Y_2}\cos\theta_2 \qquad (13)$$

Equations (10) and (13) correspond directly to equations (8) and (9), and we have the parallel result

$$r_\psi = \frac{\tilde{\psi}_r}{\tilde{\psi}_i} = \frac{Y_2 cos\theta_1 - Y_1 cos\theta_2}{Y_2 cos\theta_1 + Y_1 cos\theta_2} \quad (14)$$

$$t_\psi = \frac{\tilde{\psi}_t}{\tilde{\psi}_i} = \frac{2Y_2 cos\theta_1}{Y_2 cos\theta_1 + Y_1 cos\theta_2} \quad (15)$$

To write a unitary scattering matrix (for current conservation across the boundary), we need to obtain the reflection and transmission coefficients for current amplitudes from the reflection and transmission coefficients for the field amplitudes. This is given by (Dutta, 1995)

$$r_0^2 = \frac{(\tilde{\psi}_r^2/Y_1)cos\theta_1}{(\tilde{\psi}_i^2/Y_1)cos\theta_1} = r_\psi^2 \quad (16)$$

$$t_0^2 = \frac{(\tilde{\psi}_t^2/Y_2)cos\theta_2}{(\tilde{\psi}_i^2/Y_1)cos\theta_1} = t_\psi^2 \frac{Y_1 cos\theta_2}{Y_2 cos\theta_1} \quad (17)$$

From (14), (15), (16), (17), we have the current scattering matrix, in the basis $|R \rightarrow L>$ and $|L \rightarrow R>$ as

$$S_0 = \begin{pmatrix} t_{11} & r_{12} \\ r_{21} & t_{22} \end{pmatrix} = \frac{1}{Y_2 cos\theta_1 + Y_1 cos\theta_2} \begin{pmatrix} 2\sqrt{Y_1 Y_2 cos\theta_1 cos\theta_2} & Y_2 cos\theta_1 - Y_1 cos\theta_2 \\ Y_1 cos\theta_2 - Y_2 cos\theta_1 & 2\sqrt{Y_1 Y_2 cos\theta_1 cos\theta_2} \end{pmatrix} \quad (18)$$

which is unitary; in addition to this, the condition of $\gamma_1 sin\theta_1 = \gamma_2 sin\theta_2$ must also be satisfied by the electron wave.

### Electron wave incident on a slab (beam splitter)

We can extend the above analysis of the electron wave to two boundaries. Fig 2a shows the three layered symmetric system used to achieve the beam splitting. The different media have different effective refractive indexes, i.e. different applied potentials. Again, if $\tilde{a}_i$, $\tilde{a}_r$ and $\tilde{a}_t$ are the complex current amplitudes of the incident, reflected and transmitted wave, and if $\begin{pmatrix} t_{11} & r_{12} \\ r_{21} & t_{22} \end{pmatrix}$ is the scattering matrix for an electron wave travelling from medium I to medium II, then using the infinite reflection method, adding up all the reflected and transmitted components we can write,

$$\frac{\tilde{a}_r}{\tilde{a}_i} = r_{12} + t_{11} t_{22} r_{21} \frac{e^{-2j\theta}}{1 - r_{21}^2 e^{-2j\theta}} \quad (19)$$

$$\frac{\tilde{a}_t}{\tilde{a}_i} = t_{11} t_{22} \frac{e^{-j\theta}}{1 - r_{21}^2 e^{-2j\theta}} \quad (20)$$

Where $\theta = -\gamma_2 l$ is the phase added while travelling across the slab, $l$ being the length of the slab and $\gamma_2$ the propagation constant of the electron wave in the slab. Further, using (18), assuming that $r_{21} = -r_{12} = r_0$ and $t_{11} = t_{22} = \sqrt{1-r_0^2}$, (19) and (20) reduce to

$$r = \frac{\tilde{a}_r}{\tilde{a}_i} = -r_0 \frac{1-e^{-2j\theta}}{1-r_0^2 e^{-2j\theta}} \qquad (21)$$

$$t = \frac{\tilde{a}_t}{\tilde{a}_i} = (1-r_0^2) \frac{e^{-j\theta}}{1-r_0^2 e^{-2j\theta}} \qquad (22)$$

for $\theta = (4n-1)\pi/2$ we have $e^{-j\theta} = j$ and $e^{-2j\theta} = -1$, (21) and (22) become

$$r = \frac{-2r_0}{1+r_0^2} \qquad (23)$$

$$t = j\frac{1-r_0^2}{1+r_0^2} \qquad (24)$$

Substituting the value of $r_0$ from (16) into (23) and (24), and proceeding to write the scattering matrix for the slab, we have

$$S = \frac{1}{Y_2^2 \cos\theta_1 + Y_1^2 \cos\theta_2} \begin{pmatrix} j2Y_1Y_2\cos\theta_1\cos\theta_2 & Y_2^2\cos\theta_1 - Y_1^2\cos\theta_2 \\ Y_2^2\cos\theta_1 - Y_1^2\cos\theta_2 & j2Y_1Y_2\cos\theta_1\cos\theta_2 \end{pmatrix}$$

which is the general form of the beam splitter. As all the angle dependencies are *cosine* terms, we can drop them if we assume the incident angle to be small. Hence we reach the following approximate scattering matrix:

$$S = \frac{1}{Y_2^2 + Y_1^2} \begin{pmatrix} j2Y_1Y_2 & Y_2^2 - Y_1^2 \\ Y_2^2 - Y_1^2 & j2Y_1Y_2 \end{pmatrix} \qquad (25)$$

To obtain a 50-50 beam splitting, we equate the reflection and transmission coefficient magnitudes,

$$2Y_1Y_2 = Y_2^2 - Y_1^2 \qquad (26)$$

$$\Rightarrow Y_2^2 + Y_1^2 = \sqrt{2}\,(Y_2^2 - Y_1^2) \qquad (27)$$

Substituting the values of $Y_1$ and $Y_2$ from (12) into (26), assuming the effective electron mass to remain nearly equal in I and II, the condition for the 50-50 beam splitting is obtained as

$$(E-V)_1 - (E-V)_2 = 2\sqrt{(E-V)_2(E-V)_1} \qquad (28)$$

given (28), and using (26) and (27), the 50-50 beam splitter is obtained with the following scattering matrix,

$$S = \frac{1}{\sqrt{2}} \begin{pmatrix} j & 1 \\ 1 & j \end{pmatrix} \qquad (29)$$

The unitary beam splitter described above was experimentally demonstrated (R.C.Liu, 1998) as shown in Fig S2b; the scattering matrix derived in (29) successfully explains the antibunching of fermions and bunching of bosons as observed by Liu et al.

## Conclusion

The beam splitter matrix was obtained as $\frac{1}{\sqrt{2}} \begin{pmatrix} j & 1 \\ 1 & j \end{pmatrix}$. This scattering matrix satisfies all the conditions initially demanded of the beam splitter; namely 50-50 beam splitting and unitarity of the scattering matrix. The scattering matrix is also found to be symmetric, which is expected as the beam splitter structure is symmetric. We shall now demonstrate that the antibunching property is also satisfied. Fig S3 shows the two possible ways in fermion 1 is received at the left output and fermion 2 is received at the right output.

If $\begin{pmatrix} t_{11} & r_{12} \\ r_{21} & t_{22} \end{pmatrix}$ is the scattering matrix, then the antibunching condition is satisfied if

$$t_{11} t_{22} - r_{12} r_{21} = 1$$

which is clearly satisfied for the matrix as $\frac{1}{\sqrt{2}} \begin{pmatrix} j & 1 \\ 1 & j \end{pmatrix}$.

## B. DESCRIPTION OF THE MODEL

Two qubit quantum algorithms have been demonstrated in our device SQualD through Non Equilbrium Green's function simulations, thereby modeling spin andphase coherent transport in two dimensional electron gas. Electron spin precession being already demonstrated in InGaAs/GasAs heterostructure we take values corresponding to that system for our modeling.

Single band effective mass Hamiltonian has been used to represent the transport of electrons in conduction band. Since we consider a two dimensional electron gas (2DEG), the electrons are free to move in x and y plane but confined in z direction. Thus there wil be sub-bands in the z direction with different wave functions and cut-off energies. At low temperature only the lowest subband will be occupied. Thus we can write the effective Hamiltonian as

$$H = -\frac{\hbar^2}{2m}\left(\frac{d^2}{dx^2} + \frac{d^2}{dy^2}\right) + U(x,y) + E_s \qquad (i)$$

Where m= effective mass of electron

$E_s = E_c + \varepsilon_1$ where $E_c$ is the conduction band energy and $\varepsilon_1$ is the lowest sub-band energy in z direction

We next discretize the spatial coordinates to get the 2D tight binding Hamiltonian matrix H

$[H]_{ij}$     = $U_i + 4t$     if i=j
           = -t           if i and j are nearest neighbours
           = 0            otherwise

Where $t = \frac{\hbar^2}{2ma^2}$ ( a is the separation between the grid points represented by i and j's ). In our simulations a=1nm, m= $0.05 \times m_0$ where $m_0$ is the effective mass of the electron.

As explained in text transverse electric field is applied in the nanowires to cause spin orbit interaction by Rashba effect. So the following Rashba term is added to the Hamiltonian :

$$E = \frac{i\alpha}{2a} (\sigma_y e^{ik_x a} - \sigma_y e^{-ik_x a} - \sigma_x e^{ik_y a} + \sigma_x e^{-ik_y a}) \tag{ii}$$

Thus in two dimensional discete space we get the effective Hamiltonian as

$$H = (\sum_{i=1}^{N}(4t)c_i^+ c_i)\, I + \sum_{<i,j>}((-t)I + i\left(\frac{\alpha}{2a}\right)\sigma_y - i\sigma_x\left(\frac{\alpha}{2a}\right))\, c_i^+ c_j + h.c. \tag{iii}$$

where α is the Rashba parameter which will only be presented in the selected region where the transverse electric field is applied, I is identity matrix , $\sigma_x$ is the Pauli spin matrix in x direction and $\sigma_y$ is the Pauli spin matrix in y direction.

Green's functions are evaluated and then trasmission vs energy and current vs voltage characterstics are calculated using standard NEGF formalism

### C.IMPLEMENTATION OF DEUTSCH'S ALGORITHM

We first demonstrate the Deutsch Jotza algorithm which is a quantum algorithm that can determine whether a function mapping from N bits to 1 bit is a constant function or a balanced function in 1 function call as opposed to its classical equivalent which needs $2^{N-1} + 1$ function calls(17).This algorithm can be well illustrated from its simplest possible case – function mapping from one bit to one bit. There are four possible functions, two of which are constant -$f_0(x) = 0$ and $f_1(x) = 1$ , and two functions which are balanced $f_2(x) = x$ and $f_3(x) = 1-x$. We demonstrate through NEGF simulations that our device is capable of determining whether the function is constant or balanced from one call only as theoretically explained in the Deutsch Jotzsa algorithm. We use pseudo-spin as the main qubit and spin of electron as the ancillary qubit in this two qubit algorithm which is schematically explained in the gate sequence diagram. Since the initial state is |01> electrons are injected from input port 0 (pseudo spin state |0>) at down polarized state (spin |1>) .

The theoretical steps of Deutsch's algorithm as shown in the gate sequence diagram are implemented as follows

1. Input spin is |1>(down). Hadamard gate on spin is realized through a rotation about y-axis (electric field along z axis) by 90 degrees. Actually H=$iR_y(\pi/2)R_z(\pi)$ but since we are just operating on the basis states here , rotation about z axis is not needed.Thus |1> becomes $\frac{1}{\sqrt{2}}$( |0> - |1> )
2. The input pseudo spin is |0>. So the electrons are introduced into the system from upper nanowire( nanowire '0'). The coupling gate( tuned as Hadamard gate) works on the pseudo spin state |0> and makes it $\frac{1}{\sqrt{2}}$( |0> + |1> )
3. The oracle for Deutsch's algorithm operates on |x>|y> to give  |x>|y XOR f(x)>

So based on the four cases we have the following implementations:

Case 1: $f_1(x)$= 0
In this case |00> remains |00> , |01> remains |01>, |10> remains |10> and |11>  remains |11>

Thus unitary transformation performed by the oracle is I $\otimes$ I, hence all Rashba fields are switched off in the section of the device representing the oracle.

Case 2: $f_2(x)$=1
In this case |00> becomes |01> and |01> becomes |00>
In this case |10> becomes |11> and |11> becomes  |10>

Thus unitary transformation performed is I $\otimes$  NOT , so pauli X gate in the upper and lower channel will serve. Pauli  X gate :

$$\begin{bmatrix} 0 & 1 \\ 1 & 0 \end{bmatrix} = i\ R_y(\pi)\ R_z(\pi)$$

We found through NEGF simulations that for a 25 nm region electric field set at 1e-10eV/m rotates the spin by 180 degrees. Thus in the 50 nm region , first 25 nm electric field is applied along y at  1e-10eV/m to produce rotation about z axis and then in next 25 nm electric field is applied along z axis  at 1e-10eV/m  to produce rotation about y axis .( Electrons are moving along the x axis).Since the operation takes place in both the channels phase factor 'i ' is irrelevant and need not be incorporated

Case 3: $f_3(x)$=x
|00> is |00> , |01> is |01> , |10> becomes |11> and |11> becomes |10>

Thus here the oracle is equivalent to CNOT gate. So Pauli Z gate realized through Rashba field as in case 2 will be present only in pseudo-spin 1 ( lower channel) here , also the phase factor  i is important here because electrons traveling in the lower nanowire ( pseudo spin state '1') should have an additional phase factor  i , i.e. , it should have a π/2 phase difference with electrons in upper nanowire( pseudo spin state '0') irrespective of the spin state of the electrons. This additional phase difference can be brought about by using a potential barrier in the path of the electrons of the lower channel which can be tuned electrically from outside.

Case 4:     $f_4(x) = 1 - x$
Here the oracle is similar to case 3, but Pauli X gate is now in upper nanowire (pseudo spin state |0>). The implementation of Pauli X gate is same as depicted in case 3.

A Hadamard transform is next performed on just pseudo spin qubit as in step 2, and then we measure the transport (flow of current) by making the right electrical contact to either lower nanowire or upper nanowire at output. If the final pseudo spin state of the electrons is |0>, current will flow when right contact is made with upper nanowire (nanowire '0') while zero current flows when right contact is made with lower nanowire ( nanowire '1') . This is the case for constant functions, i.e., $f_1(x)$ and $f_2(x)$.

If the final pseudo spin state of the electrons is |1>, current will flow when right contact is made with lower nanowire (nanowire '1') while zero current flows when right contact is made with upper nanowire ( nanowire '0') . This is the case for balanced functions, i.e., $f_3(x)$ and $f_4(x)$.
Our simulation results verify this. Thus we have been able to implement Deutsch's algorithm and determine at one call whether a function is constant or balanced in SQuaLD.

**D. CONTROL OF PSEUDO-SPIN BY EXTERNAL VOLTAGE**

The spin and pseudo-spin of the electrons are the actual inputs of the system in any logical operation (quantum or classical) but since the device must have all electrical interface , spin and pseudo-spin of electron should be controlled by applying voltages externally .The Mesosopic Stern Gerlach Apparatus , which uses very similar kind of structure as ours, can be used to separate unpolarized electrons to up and down spin electrons in a particular direction[14].It has also been shown how spin can be rotated to any other direction in the Bloch sphere by application of Rashba electric field. [22] In this section we demonstrate how the mode of the electron can be controlled just by Rashba electric field.

Let the +z polarized electrons incident at pseudo-spin 0 be represented as $A e^{ikx} \begin{bmatrix} 1 \\ 0 \end{bmatrix}$

Since the first beam splitter behaves as unitary matrix $\begin{bmatrix} 1 & 1 \\ -1 & 1 \end{bmatrix}$ the electron if continues in pseudo-spin 0 ie goes to the lower channel will have no phase shift while if it is reflected in pseudo-spin 1 it will have phase shift of π. [11,12]  In pseudo-spin 1 the spin of electron is rotated in x-z plane by angle ϴ due to Rashba effect .Thus the eletron's wave function is

$$\frac{A}{\sqrt{2}} e^{i(kx+\pi)} \begin{bmatrix} \cos\left(\frac{\theta}{2}\right) \\ \sin\left(\frac{\theta}{2}\right) \end{bmatrix} \text{ in pseudo-spin 1 and } \frac{A}{\sqrt{2}} e^{ikx} \begin{bmatrix} 1 \\ 0 \end{bmatrix} \text{ in pseudo-spin 0} \qquad (2)$$

At second beam splitter again due to the unitarity of the scattering matrix $\begin{bmatrix} 1 & 1 \\ -1 & 1 \end{bmatrix}$ electron travelling incident from pseudo-spin 0 will have π phase shift on reflection to pseudo-spin 1 and 0 phase shift on transmission to pseudo-spin 0 whereas electron incident from pseudo-spin 1 will have 0 phase shift for both reflection and transmission.

Thus at output pseudo-spin 0 or detector 1 (refer to Fig 2a) electron wave function ($\Psi_0$) is given by

$$\Psi_0 = \frac{A}{\sqrt{2}} e^{i(kx+\pi)} \begin{bmatrix} \cos\left(\frac{\theta}{2}\right) \\ \sin\left(\frac{\theta}{2}\right) \end{bmatrix} + \frac{A}{\sqrt{2}} e^{ikx} \begin{bmatrix} 1 \\ 0 \end{bmatrix} \qquad (3)$$

or $\quad \Psi_0 = \sqrt{2} e^{ikx} \sin\left(\frac{\theta}{4}\right) \begin{bmatrix} \sin\left(\frac{\theta}{4}\right) \\ -\cos\left(\frac{\theta}{4}\right) \end{bmatrix} \qquad (4)$

Thus in pseudo-spin 0 transmission probability of +z polarized electrons is $2A^2 \sin^4(\theta/4)$ and of −z polarized electrons is $(A^2/2) \sin^2(\theta/2)$.

Similarly at output pseudo-spin 1 or detector 2 (refer to Fig 2a) electron wave function ($\Psi_1$) is given by

$$\Psi_1 = \frac{A}{\sqrt{2}} e^{i(kx+\pi)} \begin{bmatrix} \cos\left(\frac{\theta}{2}\right) \\ \sin\left(\frac{\theta}{2}\right) \end{bmatrix} + \frac{A}{\sqrt{2}} e^{i(kx+\pi)} \begin{bmatrix} 1 \\ 0 \end{bmatrix} \qquad (5)$$

or $\Psi_1 = \sqrt{2} A e^{i(kx+\pi)} \cos\left(\frac{\theta}{4}\right) \begin{bmatrix} \cos\left(\frac{\theta}{4}\right) \\ \sin\left(\frac{\theta}{4}\right) \end{bmatrix} \qquad (6)$

Thus in pseudo-spin 1 transmission probability of +z polarized electrons is $2A^2 \cos^4(\theta/4)$ and of −z polarized electrons is $(A^2/2) \sin^2(\theta/2)$. These analytically obtained results fully agree with the NEGF simulation results of Fig 2b. Similarly it can be shown that if −z polarized electrons are incident in pseudo-spin 0, transmission probability of −z polarized electrons in output pseudo-spin 0 is $2A^2 \sin^4(\theta/4)$ and in output pseudo-spin 1 is $2A^2 \cos^4(\theta/4)$, while transmission of +z polarized electrons in both output pseudo-spins 0 and 1 is

$(A^2/2) \sin^2(\theta/2)$. Thus when gate voltage at pseudo-spin 1 (Fig 1) is turned off i.e. Rashba electric field is 0, all electrons will be obtained at pseudo-spin 1 with no spin flip (Fig 2b) whereas when gate voltage is turned on such that Rashba electric field =+$10^{10}$ eV/m or -$10^{10}$ eV/m all electrons will be obtained at pseudo-spin 0 with no spin flip (Fig 2b). Thus we are able to control the pseudo-spin of the electron without affecting its spin by applying external voltage only.

**Supplementary Section References**

**Supplementary Section Figure Captions**

**Figure S1**  Two media separated by an interface, a.  For an EM Wave, the two media have different refractive indexes. The incident, reflected and transmitted waves are all TEM, with the electric field parallel to the interface and the magnetic field perpendicular to the electric field and the direction of propagation. The angles of incidence and reflection are equal , and related to the angle of transmission by the Snell's law. b. The analogous electron wave case, with the two media having different applied potential. The wave function  can be thought of as analogous to the electric field, and its first spatial derivative  to the magnetic field while solving the boundary conditions.

**Figure S2**  The beam splitter structure a. The electron wave is incident on a slab of medium II, with medium I on either side.   are the amplitudes of the probability current. The oblique angle of incidence allows the left input and the left output, or the right input and right output to be separated along the vertical direction.  b. The  experimental structure of the beam splitter as demonstrated by Liu et al (R.C.Liu, 1998)

**Figure S3** The antibunching of fermions as demonstrated by Liu et al. The interchanging of fermions leads to a flip in the sign of the wavefunction, which gives the antibunching of electrons at the output. This result is also expected from the Pauli Exclusion Principle.

**Figure S4**
 a- At the input port or output port electrons can be localized in either the upper or lower channel which we call the pseudo spin states, while after passing through the beam splitter they are in a superposition of these two states. The channels corresponding to the pseudo spin states |0> and |1> are marked in the figures .Blue shading indicates the path electrons follow. Since a pair of beam splitters act as an inverter and there is no Rashba field in this case, electrons incident from input port 0 (pseudo spin state |0> ) will transmit to output port 1 ( pseudo spin state |1>)  or rather the pseudo spin state is inverted. The spin of the electrons doesn't flip in the process. Thus we can initialize the electrons to pseudo spin state |1>
b- Here Rashba electric field is turned on at $1 \times 10^{-10}$eV/m .In this case as explained in the next figure electrons incident from input port 0( pseudo spin state |0> ) will transmit to output port 0 (pseudo spin state |0> ) without the spin being flipped. Thus by turning the Rashba electric field on we can initialize the electrons to pseudo spin state |0>

c-The curves represent the variation of the transmission probability of electrons incident from input port 0 with spin in +z direction ( spin up) to the output ports 0 and 1 with the applied Rashba electric field on electrons in upper channel between the two beam splitters as shown in Fig 2b. We see that when the transverse electric field causing Rashba effect is 0 eV/ m   transmission probability of spin up electrons to output port 0 is 0 and to input port 1 is maximum .Transmission probability of spin down electrons is 0 at both ports which means spin doesn't flip. This explains  Fig 2a.When transverse electric field causing Rashba effect is set at $1 \times 10^{-10}$ eV/m transmission probability of spin up electrons to output port 0 is maximum and to output port 1 is 0 and transmission probability of spin down electrons to both the output ports is 0 which  explains Fig S4 b.

**Supplementary Figures**

Fig S1

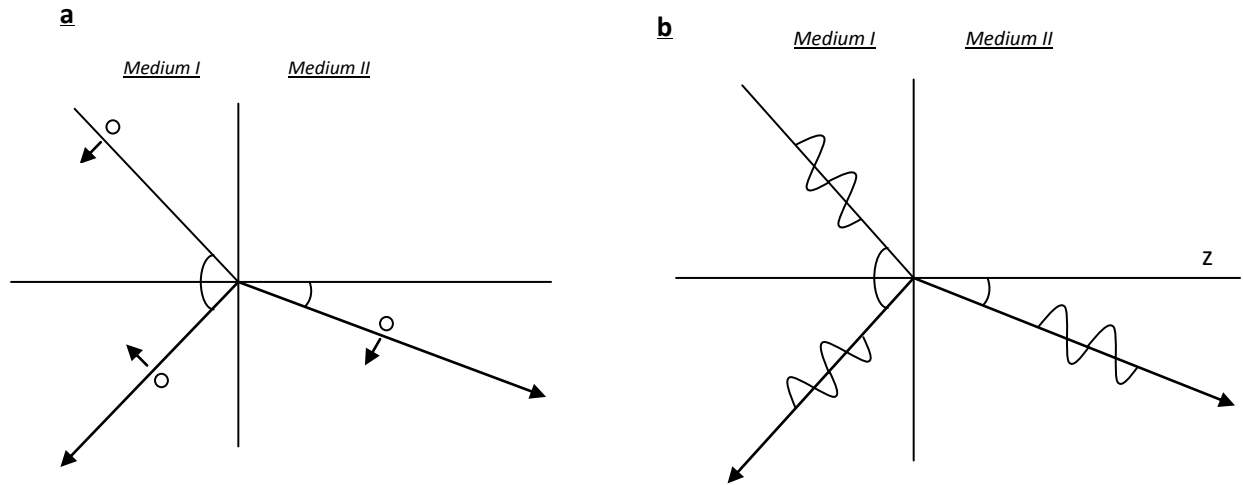

Fig S2

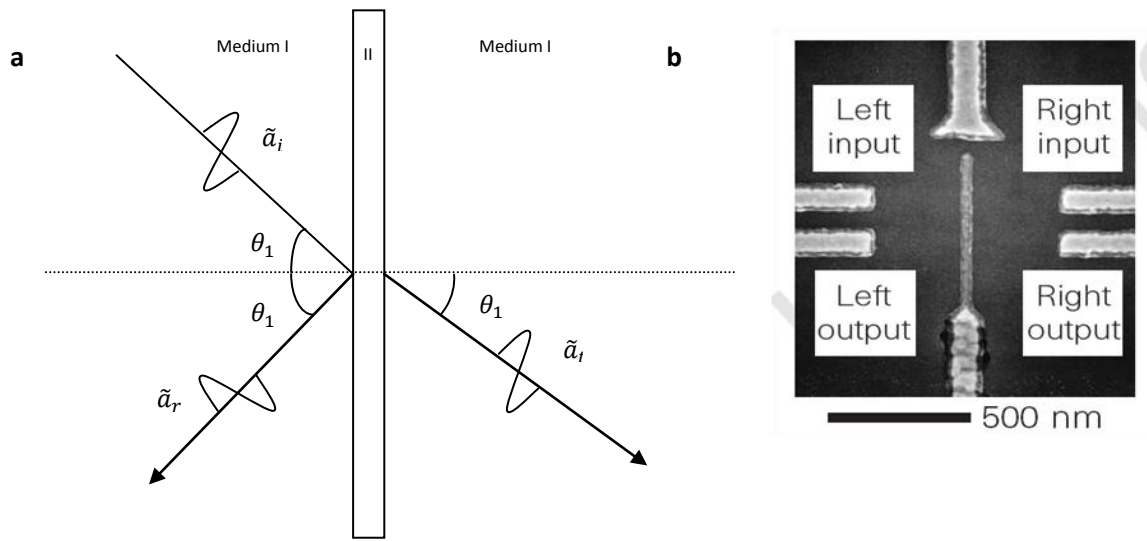

Fig S3

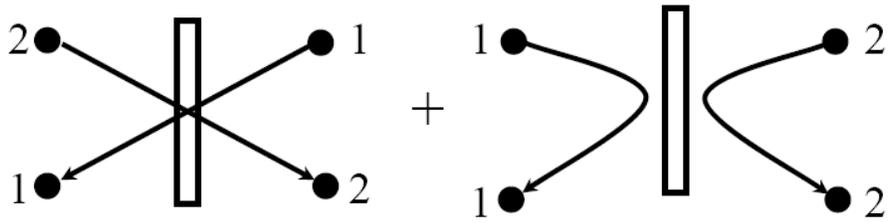

Fig S4

a

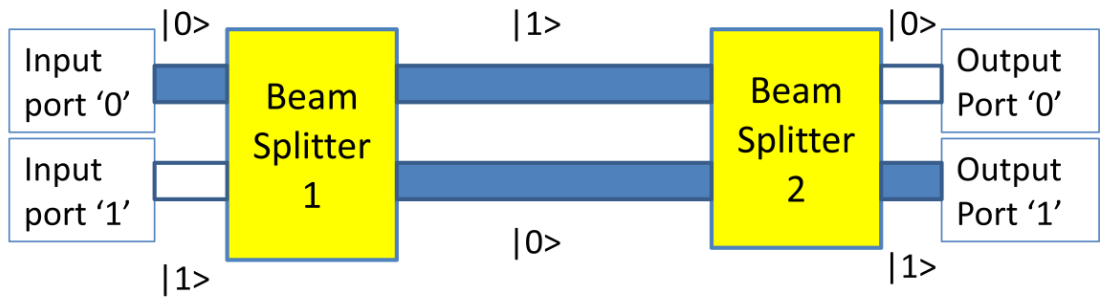

b

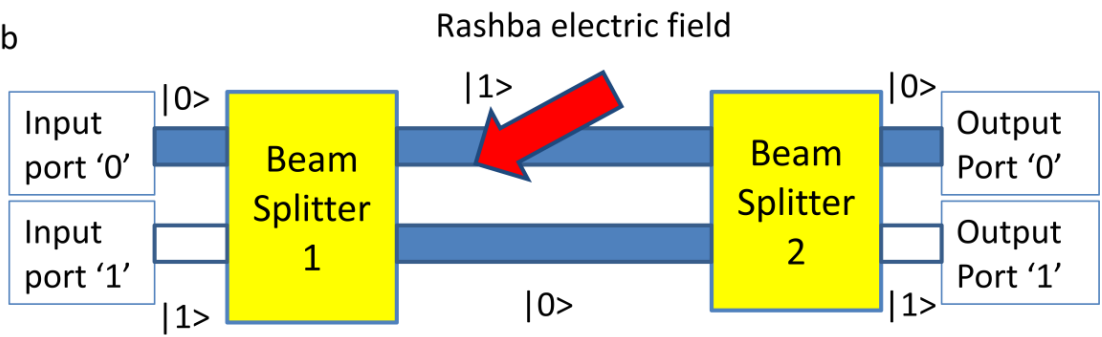

c

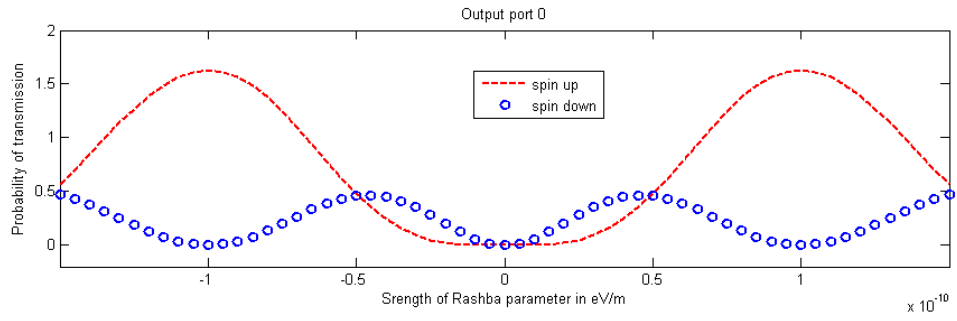

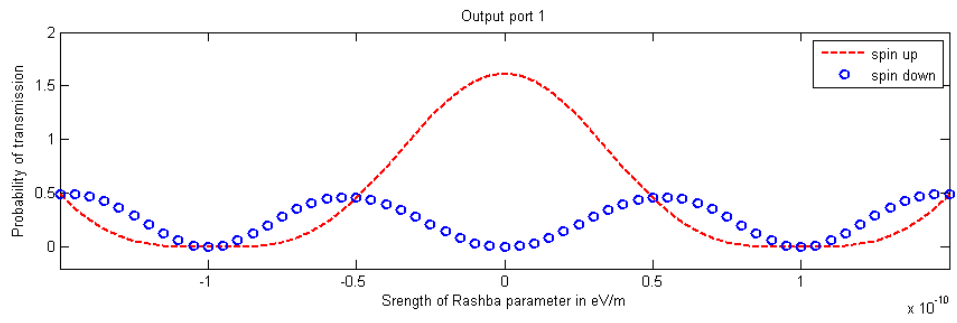